\documentclass[twocolumn,showpacs,aps,prl,superscriptaddress]{revtex4}

\usepackage{graphicx}
\usepackage{dcolumn}
\usepackage{amsmath}
\usepackage{amssymb}
\usepackage{epsfig}
\usepackage{amsgen}
\usepackage{amsmath}
\usepackage{amsgen}
\usepackage{amsmath}
\usepackage{amsfonts}

\input pubboard/babarsym

\def\eb    {\ensuremath{\varepsilon_{\it B}}}
\def\reb   {\ensuremath{\rm{Re}(\varepsilon_{\it B})}}

\def\bLH    {\ensuremath{|B^0_{L,H}\rangle}}

\def\BzBz   {\ensuremath{\Bz {\kern -0.16em \Bz}}\xspace}
\def\BzbBzb {\ensuremath{\Bzb {\kern -0.16em \Bzb}}\xspace}

\def\etpm   {\ensuremath{\varepsilon^\pm_{track}}}
\def\eppm   {\ensuremath{\varepsilon^\pm_{pid}}}
\def\nppm   {\ensuremath{\eta^\pm_{pid}}}

\def\T      {\ensuremath{T}\xspace}

\def\AT     {\ensuremath{A_{T/CP}}}

\def\aebf {asymmetric-energy \BF}

\newcommand{\BABARPubYear}    {01}
\newcommand{\BABARPubNumber}  {20}

\newcommand{\SLACPubNumber} {9149}

\def\figurebox#1#2#3{%
    \def\arg{#3}%
    \ifx\arg\empty
    {\hfill\vbox{\hsize#2\hrule\hbox to #2{\vrule\hfill\vbox to #1{\hsize#2\vfill}\vrule}\hrule}\hfill}%
    \else
    {\hfill\epsfbox{#3}\hfill}%
    \fi}

\begin{document}

\preprint{\babar-PUB-\BABARPubYear/\BABARPubNumber}
\preprint{SLAC-PUB-\SLACPubNumber}

\begin{flushleft}
\babar-PUB-\BABARPubYear/\BABARPubNumber\\
SLAC-PUB-9149
\end{flushleft}

\title{
\vskip 10mm
{\large \bf
Search for  $T$ and \CP\  Violation in \Bz -\Bzb Mixing
with Inclusive Dilepton Events}
}

%
\author{B.~Aubert}
\author{D.~Boutigny}
\author{J.-M.~Gaillard}
\author{A.~Hicheur}
\author{Y.~Karyotakis}
\author{J.~P.~Lees}
\author{P.~Robbe}
\author{V.~Tisserand}
\author{A.~Zghiche}
\affiliation{Laboratoire de Physique des Particules, F-74941 Annecy-le-Vieux, France }
\author{A.~Palano}
\author{A.~Pompili}
\affiliation{Universit\`a di Bari, Dipartimento di Fisica and INFN, I-70126 Bari, Italy }
% ...
\collaboration{The \babar\ Collaboration}
\noaffiliation

\date{\today}

\begin{abstract}
We report the results of a search for  \T and \CP violation in \Bz -\Bzb mixing using an
inclusive dilepton sample collected by the \babar\ experiment at the
\pep2\ \BF. The asymmetry between
$\ellp\ellp$ and $\ellm\ellm$ events
allows us to compare the probabilities for $\Bzb \to \Bz$ and
$\Bz \to \Bzb$ oscillations and thus probe \T and \CP invariance.
Using a sample of 23 million \BB\ pairs,
we measure a same-sign dilepton asymmetry of
$\AT=(0.5\pm1.2({\rm stat})\pm1.4({\rm syst}))\,\%$.
For the modulus of the ratio of complex mixing parameters $p$ and $q$, we obtain
$|q/p|=0.998\pm0.006({\rm stat})\pm0.007({\rm syst})$.

\end{abstract}


\maketitle

Since the first observation of \CP violation in 1964~\cite{CP64}, the kaon system
has provided many other results probing the \CPT and \T discrete
symmetries~\cite{CPLEAR99}.
Beyond the investigation of \CP violation through the
measurements of   the unitarity triangle angles $\alpha$, $\beta$ and $\gamma$,
 the \babar\ experiment can investigate
 \T and \CP violation purely in mixing.

The physical states (solutions of the complex effective Hamiltonian 
for  the \Bz -\Bzb system) can be written as
$$
\bLH = p|\Bz\rangle \pm\, q|\Bzb\rangle
$$
where $p$ and $q$ are complex mixing parameters with the normalization
$|p|^2 + |q|^2 = 1$. 

The \CPT invariant asymmetry, $\AT$, between the two oscillation probabilities
$P(\Bzb \to \Bz)$ and $P(\Bz \to \Bzb)$  probes both \T and \CP symmetries
and can be expressed in terms of $p$ and $q$:
\begin{equation}
\begin{split}
\AT& =  \frac {P(\Bzb \to \Bz)-P(\Bz \to \Bzb)}
                       {P(\Bzb \to \Bz)+P(\Bz \to \Bzb)}\\
&=   \frac {1-|q/p|^4}{1+|q/p|^4}.
\end{split}
\label{at}
\end{equation}
Standard Model calculations~\cite{SM} predict
the size of this asymmetry to be at or below $10^{-3}$. 
Therefore, a large measured value could be an
indication of new physics.

Inclusive dilepton events  representing 4\% of all \upsbb
decays  provide a very large sample with which to study \T and \CP violation in mixing.
The flavor of each  $B$ meson is tagged by the charge of the lepton.
Assuming $\Delta B=\Delta Q$ and
\CP invariance in the direct ($b \to \ell$) semileptonic decay process,
the asymmetry between same-sign lepton pairs, $\ellp \ellp$ and
$\ellm \ellm$, allows a comparison of the two oscillation probabilities
$P(\Bzb \to \Bz)$ and $P(\Bz \to \Bzb)$.
The asymmetry $\AT$ for direct same-sign dileptons is time
independent. However, in this analysis, the time difference \deltat\ between
 the two $B$ meson decays
is used to discriminate the direct leptons from the cascade leptons produced
in ($b \to c \to \ell$) transitions.

The measurement of $\AT$ reported here
is performed using events collected by  the
\babar\ detector~\cite{BaBarNIM01} 
from $e^+e^-$ collisions
at the \pep2 \aebf\ between October 1999 and October 2000.
The integrated luminosity of this sample is 20.7 \invfb recorded
at the \FourS resonance (``on-resonance'')
and 2.6 \invfb recorded about 40 \mev below the \FourS\  resonance (``off-resonance'').
\BB\ pairs from the \FourS\ decay move along the high-energy beam direction ($z$)
with a nominal Lorentz boost  $\langle \beta \gamma \rangle = 0.55$.

Lepton candidates must  have at
least 12 hits in the drift chamber (DCH), at least one $z$-coordinate
hit in the silicon vertex tracker (SVT),
and  a momentum in the \FourS center-of-mass system (CMS) between
0.7 and 2.3 \gevc.
Electrons are selected by  requirements
on the ratio of the energy deposited in the electromagnetic calorimeter (EMC)
and the momentum measured in the DCH,
on the lateral shape of the energy deposition in the calorimeter, and on the
specific ionization density measured in the DCH.
Muons are identified through the energy released in the calorimeter, as
well as the strip multiplicity, track continuity and penetration depth in the
instrumented flux return (IFR). Lepton candidates are rejected if they are
consistent with a kaon or proton hypothesis according to the Cherenkov angle
measured in the detector of internally reflected Cherenkov light (DIRC)
or to the ionization density measured in the DCH. The electron and muon selection
efficiencies are about 92\% and 75\%, with pion misidentification probabilities around
0.2\% and 3\%, respectively.

Non-\BB events are suppressed by
requiring the  ratio of second to zeroth order Fox-Wolfram moments~\cite{FW}
to be less than 0.4.
In addition, the residual contamination from radiative Bhabha and two-photon
events is
reduced by requiring the squared invariant mass of the event to be greater
than 20\,GeV$^2/c^4$, the event aplanarity to be greater than 0.01, and the
number of charged tracks to be greater than four.
Electrons from photon conversions are identified and rejected
with a negligible loss of efficiency for signal events.
Leptons from \jpsi\ and $\psi (2S)$ decays are identified by pairing
them with other oppositely-charged candidates of the same-lepton species,
selected with looser criteria. We reject the whole event if any combination
has an invariant mass within  $3.037< M(\ellp\ellm)<3.137 \gevcc$ or
$3.646< M(\ellp\ellm)<3.726 \gevcc$.
\begin{figure}[hbtp]
\begin{center}
\mbox{\includegraphics[height=10cm]{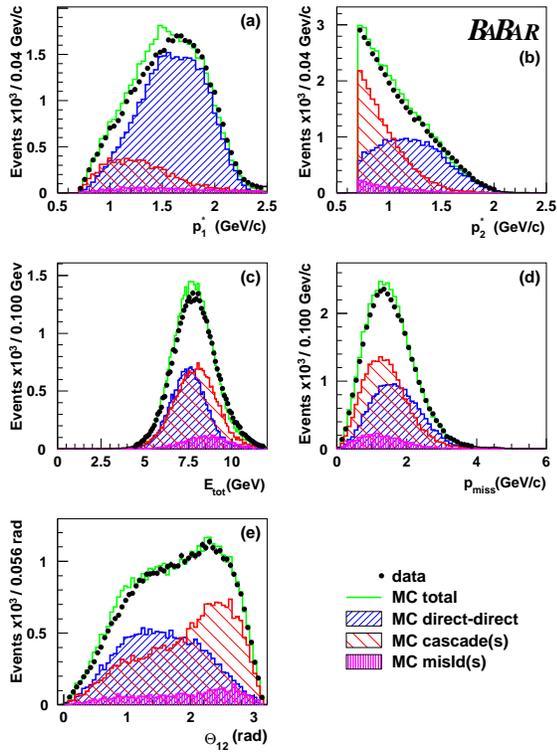}}
\end{center}
\caption{Distributions of the discriminating variables (a) $p^*_1$,
(b) $p^*_2$, (c) $E_{tot}$, (d) $p_{miss}$ and (e) $\theta_{12}$,
for data (dots) and Monte Carlo events (histograms).
The contributions from direct-direct pairs, direct-cascade or cascade-cascade pairs,
and pairs with one or more fake leptons are
shown for the Monte Carlo samples.}
\label{dis_var}
\end{figure}

To minimize the wrong flavor tags due to  leptons from cascade charm
decays, we use a neural network
algorithm that combines five discriminating variables. These are calculated
in the CMS (see Fig.~\ref{dis_var}) and are
the momenta of the two leptons with highest momentum, $p^*_1$ and $p^*_2$,
the total visible energy $E_{tot}$, the missing momentum
$p_{miss}$ of the event, and the opening angle between the leptons, $\theta_{12}$.
The first two variables, $p^*_1$ and $p^*_2$, are very powerful in
discriminating between direct and
cascade leptons. The last variable, $\theta_{12}$, efficiently removes
direct-cascade lepton pairs coming from the same $B$ and further rejects
photon conversions.
Some additional discriminating power is also provided by the other two
variables. In order to be insensitive to the Monte Carlo, the  fraction of 
cascade leptons is determined from a fit to the same-sign and opposite-sign 
dilepton data. 

\begin{figure}[hbtp]
\begin{center}
\mbox{\epsfig{file=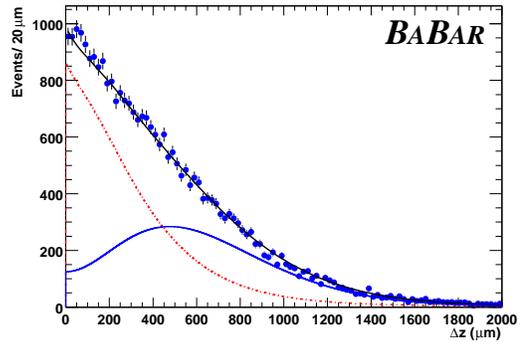,height=4.5cm}}
\end{center}
\caption{Distribution of the same-sign dileptons as a function of \deltaz;
the curve superimposed on the dots is determined from a fit to the
same-sign and opposite-sign
dileptons; the solid and dotted lines represent respectively
the signal component (\BzBz or \BzbBzb pairs)  and the background component
(cascade leptons from unmixed \BzBzb and \BpBm events,
leptons from \jpsi, resonance decays and non-\BB events).}
\label{pdfSame}
\end{figure}

In the inclusive approach used here, the $z$ coordinate of the $B$ decay point is
the $z$ position of the point of closest approach between the lepton
candidate and
an estimate of the \FourS decay point in the transverse plane. The
\FourS decay point
is obtained by fitting the two lepton tracks to a common vertex
in the transverse plane, which is constrained
to be consistent with  the beam-spot position. The time difference, $\deltat$,
between the two $B$ meson decays
is determined from the absolute value, \deltaz,  of the difference in $z$ between the two $B$
decays by
$\deltat=\deltaz/ \langle\beta\gamma\rangle c$.
The background events (cascade leptons
from unmixed \BzBzb events and \BpBm events, and non-\BB events) are most
prominent at low \deltaz (see Fig.~\ref{pdfSame}).
Therefore, a requirement of $\deltaz > 200$\mum allows us
to eliminate about 50\%
of background without dramatically decreasing the signal
efficiency.  Finally,
in the measurement of $\AT$, the dilution factor due to  remaining
background is corrected as a function of \deltat.

Application of the selection criteria described above results in a
sample of 20,381 same-sign dilepton events, consisting of 5,252
electron pairs, 5,152 muon pairs and 9,977 electron-muon pairs.
For $\deltaz > 200$\mum , the fraction of non-\BB events, measured
with the off-resonance data, is 4.3\% with a charge asymmetry of
$(-5 \pm 10)\%$; the main \BB backgrounds, determined from Monte
Carlo simulation, include 24\% of one direct lepton paired
with a cascade lepton from the other
$B$, 10\% of fake leptons from the other $B$, 2\% of fake leptons
from the same $B$ and 2\% of leptons from \jpsi\ or resonance
decays. 

Since the asymmetry $\AT$ is expected to be small, we have
carefully determined the 
possible charge asymmetries induced by the detection and
reconstruction of electrons and muons. The three sources of charge
asymmetry in the selection of lepton candidates come from 
differences, for positive and negative particles,
in tracking efficiency $\etpm$,  in
particle identification  efficiency $\eppm$, and  in
misidentification probability $\nppm$. 
Independent samples are used to estimate these efficiencies
and probabilities as a
function of several charged track parameters $x_i$: total 
or transverse momentum, and polar  and azimuthal angles
in the laboratory frame. The
numbers of ``detected'' positive and negative leptons
$N^\pm_{det}$ are related to the numbers of true leptons
$N^\pm_{true}$ by the equation
\begin{equation}
\begin{split}
N^\pm_{det}(x_i,p^*)  = &
N^\pm_{true}(x_i,p^*)\cdot\etpm(x_i)\cdot\bigl[\eppm(x_i) +\\
& r(\pi,p^*)\cdot\nppm(\pi,x_i)
 + r(K,p^*)\cdot\\
& \nppm(K,x_i) + r(p,p^*)\cdot\nppm(p,x_i)\bigr],
\end{split}
\label{weight}
\end{equation}
where $r(\pi,p^*)$, $r(K,p^*)$ and $r(p,p^*)$ are the relative
abundances of hadrons ($\pi$, $K$, and $p$) with respect to the
lepton abundance for a given $p^*$ (the momentum of the track in
the CMS). These quantities are obtained from \BB Monte Carlo
events, after applying the event selection criteria with perfect
particle identification. To correct for charge asymmetries
in the  lepton detection,
we apply a weight proportional to the ratio 
 $N^\pm_{true}(x_i,p^*)/ N^\pm_{det}(x_i,p^*)$, 
for each lepton in the sample.

Using tracks  selected from multi-hadron events, the tracking efficiencies
$\etpm(x_i)$ for positive and negative particles  are determined by computing
the ratio of the number of SVT tracks
with at least 12 DCH hits as required in the dilepton selection,
 divided by the initial number of SVT tracks.
These tracking efficiencies  are tabulated as a function of
transverse momentum and polar and azimuthal angles.
The charge asymmetry correction is less
than 0.1\% on average in the relevant momentum range.

The identification efficiencies $\eppm(x_i)$ are measured
as a function of total momentum and  polar and azimuthal angles,
with  two control samples consisting of $ee \to eeee\,({\rm with\,}\gamma\gamma\to ee)$ and
radiative Bhabha events for  electrons, and with a
 $ee \to ee\mu\mu\,({\rm with\,}\gamma\gamma\to \mu\mu) $ control sample  for  muons.
The misidentification probabilities $\nppm({\rm hadron},x_i)$ are determined using control
samples of kaons produced in $D^{*+}\to\pi^+D^0\to\pi^+K^-\pi^+$ decays
(and charge conjugate), pions produced in $K_S\to\pipi$ decays,
and one-prong and three-prong $\tau$ decays, and protons
produced in $\Lambda$ decays.

For the electrons, the charge asymmetry in the particle
identification efficiency reaches
(0.5--1.0)\% in some regions of the lepton phase space.
The impact of the charge asymmetry in misidentification is negligible
because the absolute misidentification probability for pions is extremely
small ($\sim 0.2\%$).
However, the $\Lambda$ control sample indicates a very large
misidentification
probability for antiprotons with momentum $\sim 1 \gevc$.
Such an effect is due to the annihilation
of antiprotons with nucleons in the calorimeter, which produces a signature
similar to that of
an electron.
The impact of this effect is balanced by the low relative abundance of
antiprotons in $B$ decays. Overall, antiprotons induce a charge asymmetry
of order 0.1\% and a correction is applied for this effect.

For the muons, the $ee\mu\mu$ control sample shows that the charge asymmetry
in the efficiency reaches 0.5\%. The misidentification probability for pions
 is much larger ($\sim 3\%$) than in the case of electrons but
there is no indication of
any charge asymmetry induced. On the other hand, the kaon misidentification
distribution shows
a charge asymmetry at the level of (10--20)\% due to the difference between
cross sections for \Kp and \Km meson interactions
with matter for momenta around 1 \gevc.

\begin{figure}[hbtp]
\begin{center}
\mbox{\epsfig{file=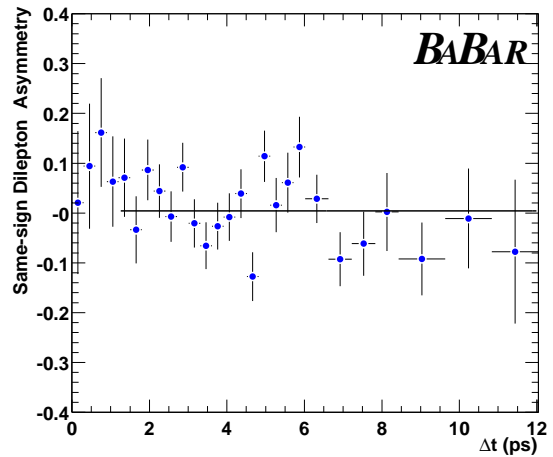,height=6cm}}
\end{center}
\caption{Corrected same-sign dilepton  asymmetry as a function of \deltat.
The line shows the result of the fit for the  dileptons with
 $\deltaz > 200$\mum.}
\label{atresultsWbckg}
\end{figure}

Equation~\ref{at} is applicable for  pure signal (direct leptons
from \BzBz and \BzbBzb events). However, the dilepton sample is
contaminated by cascade leptons from \BpBm and unmixed \BzBzb
events, non-\BB events, and \jpsi decays (see Fig.~\ref{pdfSame}).
Assuming no charge asymmetry in the background and assuming \CP
invariance holds in direct semileptonic $B$ decays, we can write
the measured asymmetry $A^{meas}_{T/CP}$ (see
Fig.~\ref{atresultsWbckg}) in terms of the number of events $N$ as
\begin{equation}
\begin{split}
A^{meas}_{T/CP}(\deltat) & =
\frac {N(\ellp\ellp,\deltat)-N(\ellm\ellm,\deltat)}
{N(\ellp\ellp,\deltat)+N(\ellm\ellm,\deltat)}\\
&= \AT\cdot \frac{S(\deltat)}{S(\deltat)+B(\deltat)},
\end{split}
\label{atmes}
\end{equation}
\noindent where $S(\deltat)$ and  $B(\deltat)$ are the numbers of signal and 
background events respectively. Therefore,
extraction of a value for $\AT$ requires a determination of the
dilution factor\break
$S(\deltat)/\left[S(\deltat)+B(\deltat)\right]$. The 
asymmetry between same-sign dileptons
is corrected for the background dilution using
the time dependent  probability density functions shown in
Fig.~\ref{pdfSame}. These probability density functions are
obtained with a fit to data for  the  same-sign and opposite-sign
dilepton samples with the value of \deltamd fixed to the world
average value~\cite{PDG}. This fit is similar to that used in the
measurement of \deltamd with dilepton events~\cite{BaBarMixing}:
it determines the corrections to the resolution function extracted
from Monte Carlo simulations, the fraction of cascade leptons, the
average lifetime of the charm component for cascade leptons, the
fraction of cascade leptons, and the fraction of charged $B$ events. In
addition,  the fraction of non-\BB events is measured from
off-resonance data. From a $\chi^2$ fit to the
distribution of the asymmetry 
as a function of \deltat\
for the same-sign dileptons with  $\deltaz > 200$\mum
(see Fig.~\ref{atresultsWbckg}), we extract $\AT=(0.5\pm1.2)\,\%$.

The systematic uncertainties related to the detection charge
asymmetry both for tracking and lepton identification  are
determined using direct leptons from semileptonic $B$ decays.
This sample  has the same topology and kinematics 
as the leptons from dilepton events. This single-lepton
charge asymmetry, sensitive to the charge asymmetry due to
detection bias,  may also be affected by the real physical 
asymmetry $\AT$ in the dilepton events. But, in practice, the
effect introduced by $\AT$ is suppressed by more than one order of
magnitude and is therefore neglected. With the 1999--2000 data
set, we select roughly 1.5 million electrons and 1.5 million
muons. After subtraction of scaled off-resonance data and after
applying a correction weight derived from Eq.~\ref{weight}, we
measure the charge asymmetries to be $(-0.30 \pm 0.14)\%$ for the
electrons and $(-0.35 \pm 0.17)\%$ for the muons. We assign these
residual asymmetries $\pm0.30\%$ and  $\pm0.35\%$ as 
systematic errors due to charge asymmetry in detection
efficiencies. With the dilution factor correction, the total
systematic errors related to the charge asymmetry in the detection
are $\pm0.5\%$ and $\pm0.6\%$ for electrons and muons,
respectively.

The assumption of no charge asymmetry in the background is confirmed by the
off-resonance data where the charge asymmetry $(-5\pm10)\%$
is consistent with zero and leads to a $\pm0.7\%$ uncertainty on the $\AT$ measurement.
In addition, the charge asymmetry of the events with $\deltaz< 100$\mum, which contain
85\% background (cascade leptons from \Bpm and unmixed \Bz),  is
$(1.2 \pm 1.4)\%$, also consistent with zero. From this asymmetry,
we can constrain to $\pm 0.9\%$ the uncertainty on $\AT$ due to a possible
charge asymmetry in the decays producing the cascade leptons. If we
assume  \CP invariance in the decays producing the cascade, this uncertainty vanishes.

The background dilution correction is measured
with the data  from the full dilepton sample with the
value of \deltamd fixed  to the world average value~\cite{PDG}. The uncertainty on the
ratio $B/S$ leads to a $\pm3\%$ multiplicative error on $\AT$, which is negligible.
A possible dilution of $\AT$ due
to double mistag is neglected because the probability of double mistag is at
the level of only 1\%.
\begin{table} [htb]
\begin{center}
\caption{Summary of  systematic uncertainties on $\AT$.}
\begin{tabular}{lc}
\hline \hline
 {\bf Type of systematic error } & \boldmath
$\sigma(\AT) (\%)$ \\
 \hline Electron charge asymmetry in the
detection & 0.5 \\ Muon charge asymmetry in the detection & 0.6 \\
Non-\BB background charge asymmetry & 0.7 \\
\BB background charge asymmetry & 0.9 \\ 
Correction of the background dilution  & 0.01 \\ 
\hline 
Total & 1.4 \\
\hline
\hline
\end{tabular}
\label{sys_table_AT}
\end{center}
\end{table}

In conclusion, we measure $\AT=(0.5\pm1.2({\rm stat})\pm1.4({\rm syst}))\,\%$ where the
total systematic uncertainty is the quadratic sum of the  systematic
uncertainties  listed in Table~\ref{sys_table_AT}.
From Eq.~\ref{at}, the $\AT$ asymmetry gives the modulus of the ratio
of complex mixing parameters $p$ and $q$ equal to
$$
|q/p|=0.998\pm0.006({\rm stat})\pm0.007({\rm syst}).
$$
This measurement can be translated into a measurement of the \CP violating parameter 
$\eb=(p-q)/(p+q)$. We obtain 
$\reb/(1+|\eb|^2)=(1.2\pm2.9({\rm stat})\pm3.6({\rm syst})) \times 10^{-3} $,
which is  the most stringent test of \T and \CP violation in
\Bz -\Bzb mixing to date and is consistent with previous
measurements~\cite{others}.

We are grateful for the excellent luminosity and machine conditions
provided by our \pep2\ colleagues, 
and for the substantial dedicated effort from
the computing organizations that support \babar.
The collaborating institutions wish to thank 
SLAC for its support and kind hospitality. 
This work is supported by
DOE
and NSF (USA),
NSERC (Canada),
IHEP (China),
CEA and
CNRS-IN2P3
(France),
BMBF
(Germany),
INFN (Italy),
NFR (Norway),
MIST (Russia), and
PPARC (United Kingdom). 
Individuals have received support from the 
A.~P.~Sloan Foundation, 
Research Corporation,
and Alexander von Humboldt Foundation.

\end{document}